\documentclass[preprint,showpacs,preprintnumbers,amsmath,amssymb]{revtex4-1}
 \usepackage{graphicx}
 \usepackage{dcolumn}
 \usepackage{bm}
 \usepackage{ulem}
 \usepackage{enumitem}
 \usepackage{slashed}
\def\q{\bm{q}}
\def\p{\bm{p}}
\def\k{\bm{k}}
\def\v{\bm{v}}

\def\r{\bm{r}}

\def\vec\epsilon{\bm{\epsilon}}

\def\q{\bm{q}}
\def\p{\bm{p}}
\def\k{\bm{k}}
\def\v{\bm{v}}

\def\r{\bm{r}}

\usepackage{xcolor}

\begin{document}

\title{Transport coefficients of heavy quarkonia comparing with heavy quark coefficients}
\author{Juhee Hong}
\affiliation{Department of Physics and Institute of Physics and Applied Physics, Yonsei University,
Seoul 03722, Korea}
\author{Su Houng Lee}
\affiliation{Department of Physics and Institute of Physics and Applied Physics, Yonsei University,
Seoul 03722, Korea}
\date{\today}

\begin{abstract}
We revisit the transport coefficients of heavy quarkonia moving in 
high-temperature QCD plasmas. 
The thermal width and mass shift for heavy quarkonia are closely related to 
the momentum diffusion coefficient and its dispersive counterpart for heavy 
quarks, respectively. 
For quarkonium at rest in plasmas the longitudinal gluon part of the 
color-singlet self-energy diagram is sufficient to determine the leading-order 
thermal width, whereas the momentum dependence is obtained from the 
transverse gluon channel. 
Using the quarkonium-gluon effective vertex based on the 
dipole interaction of color charges, we discuss the damping rate, the 
effective rest and kinetic mass shifts of slowly moving quarkonia and 
compare with the corresponding coefficients of heavy quarks.  
\end{abstract}

\maketitle

\section{Introduction}

The suppression of quarkonium production has been predicted to indicate the  
formation of quark-gluon plasmas in relativistic heavy-ion collisions 
\cite{Matsui:1986dk}. 
In a hot medium, color-screening prevents heavy quark-antiquark pairs from 
binding and bound states are dissociated by medium interactions. 
The dynamics of heavy quarkonia are characterized by transport coefficients 
such as the thermal width and mass shift.  
These two coefficients correspond to thermal corrections to the imaginary and 
real parts of the static potential for a heavy quark-antiquark pair, 
respectively \cite{Laine:2006ns,Brambilla:2008cx}.

In high-temperature plasmas, dissociation of a bound state such as 
$\Upsilon(1S)$ is described mostly by inelastic parton scattering 
\cite{Hong:2018vgp}. 
At leading order in a weak coupling expansion, the thermal width is related 
to the momentum diffusion coefficient of heavy quark 
\cite{Brambilla:2013dpa,Brambilla:2016wgg}. 
Medium interactions of heavy quarks are described by the diffusion 
coefficient and its dispersive counterpart \cite{Brambilla:2019tpt}. 
In this work, we investigate the transport coefficients of heavy 
quarkonia, especially bound states with finite spatial momentum with respect 
to plasmas. 
Although the transport coefficients for heavy quarks and quarkonia can be 
extracted from lattice QCD computations, current lattice data have large 
uncertainties and are not practical for measuring the momentum dependence of 
the heavy quark-antiquark potential. 
Furthermore, it would be useful to identify the main partonic processes 
contributing to the transport coefficients.

In previous studies, the momentum dependence of the thermal width has been 
considered phenomenologically in a moving medium because most theoretical 
investigations have been carried out in the rest frame of a bound state.  
We have computed the thermal width by calculating the dissociation cross 
section in the quarkonium rest frame and then convoluting parton distribution 
functions given by 
$f(K)=(e^{K\cdot V/T}\pm 1)^{-1}$, where 
$V=\frac{1}{\sqrt{1-v^2}}(1,\v)$ \cite{Hong:2019ade}. 
An effective field theory approach has also been employed to calculate the 
thermal width \cite{Escobedo:2013tca}. 
In the kinematical regime $T\gg \frac{1}{r}\gg m_D\gg E$, 
they have taken into account dynamic screening \cite{Weldon:1982aq,Chu:1988wh} 
(which depends on velocity) but ignored the energy transfer between a 
bound state and plasma. 
The energy transfer can be neglected at leading order for quarkonium at rest, 
but both energy and momentum transfer are required to evaluate the thermal 
width of bound states moving in quark-gluon plasmas.

To understand dynamic properties of heavy particles in hot QCD plasmas, 
there have been many 
investigations and they need to be brought together to study transport 
processes of heavy quarkonia.  
The goal of this paper is to compute the transport coefficients for heavy 
quarkonium depending on its velocity and to explain the difference 
from the heavy quark coefficients which have been investigated in literature. 
Our analysis is based on the leading quarkonium-gluon interaction 
\cite{Oh:2001rm} which has been derived from the Bethe-Salpeter amplitude 
using the dipole interaction of color charges \cite{Bhanot:1979vb}. 
In Section \ref{inelscatt}, we calculate the velocity dependence of the 
thermal width of $\mathcal{O}(g^4T^3r^2)$ through inelastic parton scattering. 
In Section \ref{dm}, the rest and kinetic mass shifts for slowly moving 
quarkonia are determined at leading order $\mathcal{O}(g^4T^3r^2)$ and 
next-to-leading order $\mathcal{O}(g^5T^3r^2)$. 
In Section \ref{damp}, we revisit the damping rate of nonrelativistic 
quarkonium by using the spectral density derived from a Schr\"{o}dinger-type 
equation.  
Finally, we summarize our result in Section \ref{summary}. 
The details on the quarkonium-gluon effective interaction are presented in 
Appendix \ref{gluodiss}.

\section{Thermal width}
\label{inelscatt}

In high-temperature QCD plasmas, quarkonium dissociation occurs through 
inelastic parton scattering 
($\Upsilon+p\rightarrow b+\bar{b}+p$, $p=g,q,\bar{q}$) 
exchanging spacelike gluons.  
The cross section calculated in hard-thermal-loop (HTL) perturbation theory 
has been shown to agree with the result obtained by potential nonrelativistic 
QCD (pNRQCD) in the kinematical regime 
$\frac{1}{r} \gg T\gg m_D\gg E$ (where $r$ is 
the distance between $b$ and $\bar{b}$, and $E$ is the binding energy) 
\cite{Hong:2018vgp}. 
The leading-order dissociation is described by taking the dipole 
interaction of color charges (which is valid at short $r$)  
and ignoring the interaction between $b$ and $\bar{b}$ after breakup in the 
large $N_c$ limit \cite{Bhanot:1979vb}. 
In the same approach, we consider the thermal width for moving quarkonium and 
compare with the momentum diffusion coefficient of heavy quark.

A bound state of heavy quarks and its interaction with partons can be 
described by the Bethe-Salpeter amplitude in the rest frame of 
quarkonium (see Appendix A) \cite{Oh:2001rm}. 
Using the effective vertex Eq. (\ref{effver}), quarkonium dissociation 
through inelastic parton scattering has been calculated \cite{Hong:2018vgp}:
\begin{equation}
\label{mtx}
|\mathcal{M}|^2
=Cg^4m^2 m_\Upsilon\, |\nabla\psi(\p)|^2 \left[\k^2
\frac{k_{10}^2}
{(\k^2+m_D^2)^2}
\bigg\{
\begin{tabular}{cc}
$(1+\cos\theta_{k_1k_2})$ & \quad ($q,\bar{q}$) \\
$(1+\cos^2\theta_{k_1k_2})$ & \quad ($g$) 
\end{tabular}
\bigg. \right] \, ,
\end{equation}
where $C=16\frac{(N_c^2-1)N_f}{N_c}$, $16(N_c^2-1)$ for $(q,\bar{q})$, $(g)$, 
respectively, 
$\psi(\p)$ is the normalized wave function for a bound state with the 
relative momentum, $\p=\frac{1}{2}(\p_1-\p_2)$, 
and $\k=\k_1-\k_2$ is the momentum transfer. 
The matrix elements are similar to those for heavy quark diffusion, 
$t$-channel gluon exchange ($b+p\rightarrow b+p$) weighted by the squared 
momentum transfer \cite{Moore:2004tg}. 
The thermal width of quarkonium is obtained by the following phase space 
integrations \cite{Hong:2018vgp}: 
\begin{multline}
\Gamma
=\frac{1}{2d_\Upsilon q^0}
\int\frac{d^3\k_1}{(2\pi)^32k_{10}}
\int\frac{d^3\k_2}{(2\pi)^32k_{20}}
\int\frac{d^3\p_1}{(2\pi)^32p_{10}}
\int\frac{d^3\p_2}{(2\pi)^32p_{20}} \\
\times (2\pi)^4\delta^4(Q+K_1-K_2-P_1-P_2)
|\mathcal{M}|^2n(k_1)[1\pm n(k_2)] \, ,
\end{multline}
where $d_\Upsilon$ is the degeneracy factor of quarkonium. 
For quarkonium at rest, the interaction rate is proportional to 
the momentum diffusion coefficient of heavy quark: 
$\Gamma^{\rm 1S}=3a_0^2\kappa$ for 
a Coulombic bound state \cite{Brambilla:2013dpa,Brambilla:2016wgg}. 
Note that Eq. (\ref{mtx}) involves only the longitudinal part of the effective 
gluon propagator as the leading contribution.  
Since the transverse gluon propagator couples with the energy 
transfer rather than the momentum transfer, the thermal width 
of quarkonium with finite momentum will be different from the momentum 
diffusion coefficient of heavy quark.

To consider the relative motion between quarkonium and thermal medium,   
we use $\cos\theta_{kk_1}=\frac{\omega}{k}+\frac{k^2-\omega^2}{2kk_1}$.  
Then the square bracket in Eq. (\ref{mtx}) becomes 
\begin{equation}
\label{mtxfactor}
\left[\frac{k^2}{2}[4k_1(k_1-\omega)-(k^2-\omega^2)]\, |D_L(K)|^2
+\frac{\omega^2}{2k^2}[(2k_1-\omega)^2+k^2](k^2-\omega^2) \, |D_T(K)|^2
\right] \, 
\end{equation}
at leading order. 
Due to the long-range interactions mediated by gluon, infrared 
divergences arise and HTL resummation is necessary. 
In Coulomb gauge, the effective propagators for soft gluon are 
\cite{Braaten:1989mz} 
\begin{eqnarray}
D_L(K)&=&\frac{1}{k^2+\Pi_L(K)} \, , \nonumber\\
D_T(K)&=&-\frac{1}{k^2-w^2+\Pi_T(K)} \, ,
\end{eqnarray}
with HTL corrections 
\cite{Weldon:1982aq,Weldon:1982bn}
\begin{eqnarray}
\label{htl}
\Pi_L(K)&=&m_D^2\left[1-\frac{\omega}{2k}\left(\ln
\frac{k+\omega}{k-\omega}-i\pi\right)\right] \, ,
\nonumber\\
\Pi_T(K)&=&m_D^2\left[\frac{\omega^2}{2k^2}+\frac{\omega(k^2-\omega^2)}
{4k^3}\left(\ln\frac{k+\omega}{k-\omega}-i\pi\right)\right] \, .
\end{eqnarray}
The leading-log thermal width is determined by a method similar to 
that developed for calculating the screening effect 
\cite{Braaten:1991dd,Braaten:1991jj,Braaten:1991we}. 
The effective propagators screen the static Coulomb interaction, eliminating 
infrared divergences. 
Although the purely static magnetic interaction is not screened, there is 
dynamic screening at nonzero frequency \cite{Weldon:1982aq}: this dynamical 
screening cuts off infrared divergences if divergences are 
only logarithmic. 
After angular integrations, the thermal width is given by  
\begin{eqnarray}
\Gamma(v)=
\frac{1}{16(2\pi)^5d_\Upsilon q_0v}\int dk_1 \int dk \int_{-vk}^{vk}d\omega 
 \int dp \, \frac{p^2}{p_{10}^2}
\, |\mathcal{M}|^2 \,
n(k_1)[1\pm n(k_1-\omega)] \, .
\end{eqnarray}
For a Coulombic bound state, we have  
$\langle r^2\rangle=\int\frac{d^3\p}{(2\pi)^3}|\nabla\psi_{1S}(\p)|^2=3a_0^2$  
and  
\begin{eqnarray}
\label{G1S}
\Gamma^{\rm 1S}(v)&=&
\frac{3Cg^4a_0^2}{128\pi^3d_\Upsilon v}
\int dk_1 \, k_1^2\, n(k_1)[1\pm n(k_1)]
\nonumber\\
&&\qquad\times
\int dk \int_{-vk}^{vk}d\omega  
\left[k^2|D_L(K)|^2+
\frac{\omega^2(k^2-\omega^2)}{k^2}|D_T(K)|^2\right] \, ,
\nonumber\\
&\simeq&\frac{(N_c^2-1)g^4T^3a_0^2}{8\pi d_\Upsilon v}
\left(1+\frac{N_f}{2N_c}\right)
\ln\frac{1+v}{1-v} \, 
\ln\frac{T}{m_D} 
\, . 
\end{eqnarray}
We notice that the velocity-dependence comes from the 
transverse gluon part: in Eq. (\ref{mtxfactor}) $\mathcal{O}(v^2)$ terms 
from the longitudinal part have been ignored compared to the leading-log term.

\begin{figure}
\includegraphics[width=0.45\textwidth]{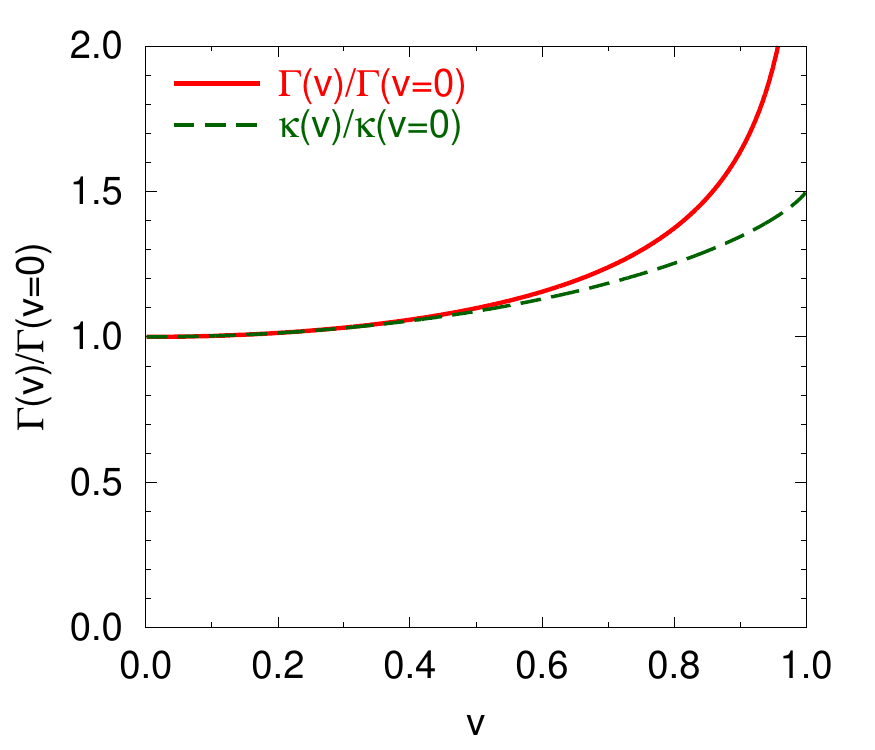}
\caption{
The velocity dependence in a leading-log approximation:  
comparing the thermal width of quarkonium and the momentum 
diffusion coefficient of heavy quark. 
}
\label{vdep}
\end{figure}

The momentum diffusion coefficient of heavy quark has been computed 
in a leading-log approximation \cite{Moore:2004tg}
\begin{eqnarray}
\kappa(v)=\frac{C_Fg^4T^3}{12\pi}\left(\frac{N_f}{2}+N_c\right)
\left[1-\frac{1-v^2}{6v}\ln\left(\frac{1+v}{1-v}\right)\right]
\ln\left(\frac{T}{m_D}\right) \, .
\end{eqnarray}
In Fig. \ref{vdep}, we present the velocity dependence of 
the thermal width, comparing with that of the heavy quark 
diffusion coefficient. 
The width increases with quarkonium momentum, in qualitative agreement with 
the previous investigation in Ref. \cite{Hong:2019ade}. 
For small $v$, $\frac{\Gamma(v)}{\Gamma(v=0)}=\frac{\kappa(v)}{\kappa(v=0)}
=1+\frac{v^2}{3}+\mathcal{O}(v^4)$: the thermal width of moving quarkonium 
agrees with the momentum diffusion coefficient of heavy quark (multiplied by 
$\langle r^2\rangle$) up to $\mathcal{O}(v^2)$.

\begin{figure}
\includegraphics[width=0.40\textwidth]{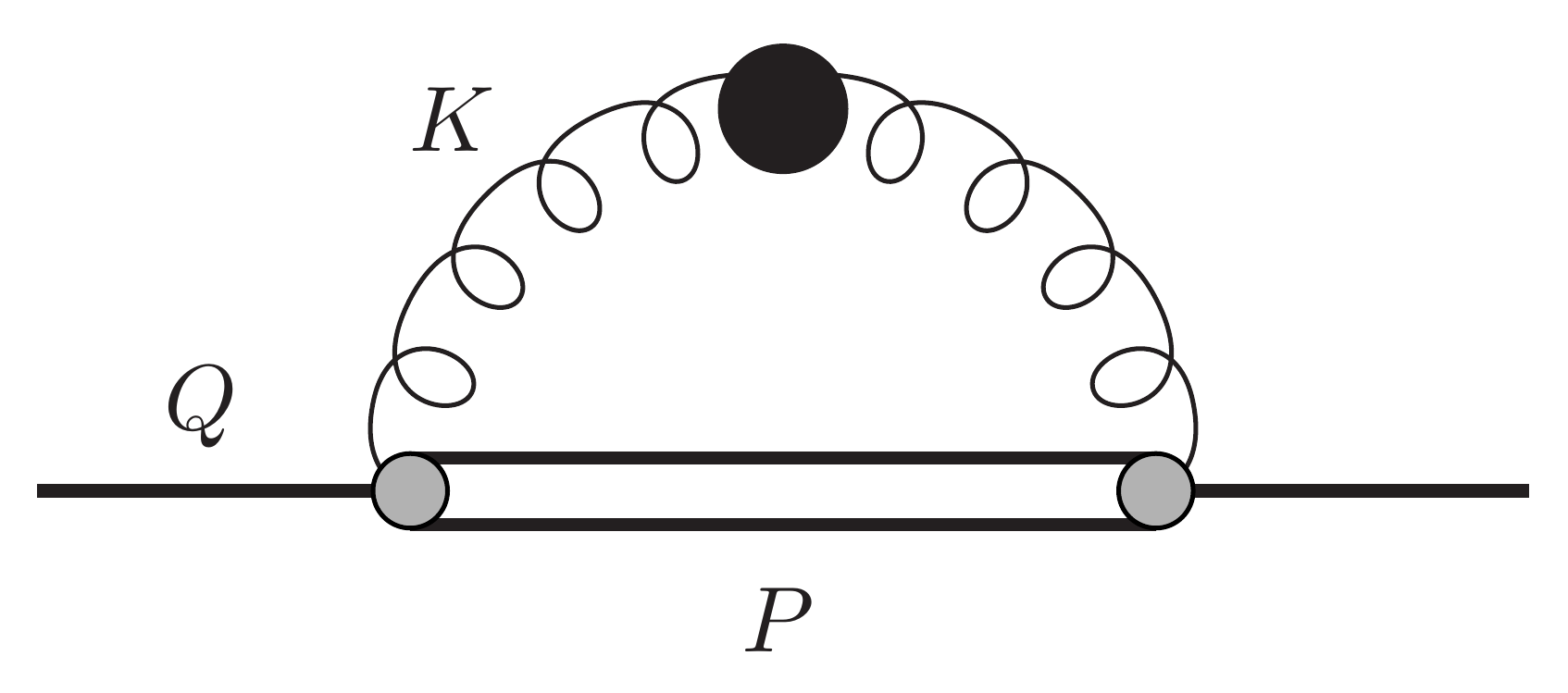}
\caption{
The self-energy diagram for heavy quark-antiquark color-singlet with 
the effective gluon propagator.}
\label{YselfE}
\end{figure}

Dynamics of heavy quarks and quarkonia is characterized by dissipative 
transport coefficients and their dispersive counterparts.
To determine the former such as heavy quark diffusion and damping rate, we 
consider the interaction rate of heavy particles. 
The interaction rate is proportional to the imaginary part of the 
self-energy and is related to the spectral density of soft gluon 
\cite{Weldon:1983jn,Braaten:1991jj}. 
On the other hand, dispersive coefficients such as mass shift corresponds to 
the real part of the gluon self-energy. 
Indeed, the thermal width and mass shift of quarkonia are given by the 
imaginary and real parts of the color-singlet self-energy  
(see Fig. \ref{YselfE}) \cite{Brambilla:2008cx,Brambilla:2016wgg}.

We reexpress the thermal width in terms of the spectral density of 
soft gluon, 
$\rho_{L,T}(K)=-\frac{1}{\pi}\,{\rm Im}\,D_{L,T}(\omega+i\epsilon,\k)$:  
\begin{multline}
\label{Gamma-Pi}
\Gamma^{\rm 1S}(v)=
\frac{3\pi(N_c^2-1) g^2a_0^2}{N_cd_\Upsilon }
\int \frac{d^3\k}{(2\pi)^3} \int d\omega \, 
[1+n_B(\omega)] \,  \delta(\omega-\k\cdot\v)
\\
\times\Big[k^2 \, \rho_L(K) 
+2\omega^2  \, \rho_T(K) \Big] \, , 
\end{multline}
where quark and gluon contributions correspond to  
$m_{D(q,g)}^2=\frac{2N_{f,c}g^2}{T}\int\frac{d^3k_1}{(2\pi)^3} \, n(k_1)[1\pm n(k_1)]$ 
in HTL corrections. 
For $\omega\ll T$, the bosonic thermal distribution is expanded as 
$1+n_B(\omega)=\frac{T}{\omega}+\frac{1}{2}+\cdots$. 
Since the spectral density is odd in $\omega$, the first term, 
$\frac{T}{\omega}$, contributes to the thermal width. 
This limit is consistent with the soft momentum transfer in the thermal width 
in Eq. (\ref{G1S}), and $\Gamma^{\rm 1S}$ for small $v$ will be reproduced 
later when discussing the damping rate.   
Replacing the spectral density by the corresponding real part in 
Eq. (\ref{Gamma-Pi}), 
the second term, $\frac{1}{2}$, produces the thermal mass 
shift which will be evaluated in the next section.

\section{Mass shift}
\label{dm}

The thermal width $\Gamma\sim g^4T^3a_0^2$ is determined through 
inelastic parton scattering with soft momentum transfer by using HTL resummed 
perturbation theory.   
The thermal mass shift is considered in the same way, 
but the leading contribution $\delta m_{\rm LO}^{\rm 1S}
=-\frac{1}{4\pi^2}\zeta(3)\left(N_f+\frac{4}{3}N_c\right)g^4T^3a_0^2$ 
comes from momenta larger than $\mathcal{O}(gT)$ scale 
\cite{Brambilla:2008cx}. 
It turns out that soft momentum transfer yields 
$\delta m_{\rm NLO}\sim g^5T^3a_0^2$ which is an $\mathcal{O}(g)$ higher term: 
the higher order corrections for the heavy quark diffusion coefficient are 
known to be significant at realistic values of the strong coupling constant 
\cite{Caron-Huot:2007rwy}. 
The mass shift provides thermal corrections to the dispersion relation of 
heavy particles, $E_{\p}=m_{\rm rest}+\frac{\p^2}{2m_{\rm kin}}$, where 
$m_{\rm rest}$ and $m_{\rm kin}$ are the rest mass and the kinetic mass, 
respectively \cite{Chesler:2009yg}. 
For $m\gg T$, the leading corrections to the mass-shell 
seem to be independent of the spatial momentum. 
In this section, we discuss the rest mass shift (which has been computed in 
Ref. \cite{Brambilla:2008cx}) and the kinetic mass shifts for slowly moving 
quarkonia, focusing on smaller momentum transfer than incoming parton momentum.

The dispersive counterpart of Eq. (\ref{Gamma-Pi}) is given by 
\begin{eqnarray}
\delta m_{\rm NLO}^{\rm 1S}(v)&=&
\frac{3(N_c^2-1)g^2a_0^2}{2N_cd_\Upsilon}
\int \frac{d^3\k}{(2\pi)^3} \int d\omega \, 
\frac{1}{2} \, \delta(\omega-\k\cdot\v) \nonumber\\
&&\qquad\times \, {\rm Re}
\left[\frac{k^2}{k^2+\Pi_L(K)}
-\frac{2\omega^2}{k^2-\omega^2+\Pi_T(K)}\right] \, .
\end{eqnarray}
For small $v$, we expand the longitudinal gluon propagator:  
\begin{eqnarray}
\label{dmY}
\delta m_{\rm NLO}^{\rm 1S}(v)&=&
\frac{3(N_c^2-1)g^2a_0^2}{2N_cd_\Upsilon }
\int \frac{d^3\k}{(2\pi)^3} \int d\omega \, 
\frac{1}{2} \, \delta(\omega-\k\cdot\v) \nonumber\\
&&\qquad\times \, 
k^2\left[
\frac{1}{k^2+m_D^2}+\frac{m_D^2\omega^2}{(k^2+m_D^2)^3}
+\frac{(4-\pi^2)m_D^4\omega^2}{4k^2(k^2+m_D^2)^3}
+\cdots\right] \, ,
\nonumber\\
&=&\frac{3(N_c^2-1)g^2m_D^3a_0^2}{16\pi N_cd_\Upsilon}\left[1
-\frac{1}{2}\left(1+\frac{\pi^2}{16}\right)v^2\right]
+\mathcal{O}(v^4) \, ,
\end{eqnarray}
which has been evaluated in dimensional regularization. 
Subtracting from the integrand the corresponding unresummed propagator 
($\frac{1}{k^2-\omega^2+\Pi_T(K)}-\frac{1}{k^2-\omega^2}$), we can show that 
the transverse gluon contributes to $\mathcal{O}(v^4)$. 
The first and second terms in the last line of Eq. (\ref{dmY}) correspond to 
the rest mass shift and the kinetic mass shift, respectively. 
We note that the effective kinetic mass shift is not equal to the rest mass 
shift. 
Similarly, the dispersion relation of heavy quark has been discussed in 
Ref. \cite{Chesler:2009yg}:
\begin{equation}
\delta m_{HQ}=
-\frac{C_Fg^2m_D}{8\pi}+\frac{1}{2}\left(1-\frac{\pi^2}{16}\right)
\frac{C_Fg^2m_D}{24\pi}v^2 \, .
\end{equation}
In comparison to heavy quark $\delta m_{HQ}\sim g^3T$, the thermal mass shift 
of quarkonium is suppressed by a factor $\sim (m_Da_0)^2$ in the case of soft 
momentum transfer.

The leading-order mass shift of moving quarkonia is  
\begin{multline}
\delta m_{\rm LO}^{\rm 1S}(v)
=-\frac{3(N_c^2-1)g^2a_0^2}{2N_cd_\Upsilon}\int\frac{d^3\k}{(2\pi)^3}
\int d\omega \,
\frac{1}{2}\, \delta(\omega-\k\cdot\v)
\left[\frac{{\rm Re}\,\Pi_L(K)}{k^2}+
\frac{2\omega^2{\rm Re}\,\Pi_T}{(k^2-\omega^2)^2}\right] \, .
\end{multline}
Here, we expand the gluon self-energy 
\cite{Kajantie:1982xx,Heinz:1986kz,Kapusta:2006pm} in $\omega$:
\begin{eqnarray}
\label{PiL}
\Pi_L(K)&=&
\frac{g^2N_f}{2\pi^2}\int_0^\infty dk_1 \, k_1n_F(k_1)\left[2+
\left(\frac{2k_1}{k}-\frac{k}{2k_1}\right)\ln\frac{k+2k_1}{k-2k_1}
\right.\nonumber\\
&&\qquad\left.
-\frac{\omega^2}{k^2}\left(\frac{2(k^2-8k_1^2)}{k^2-4k_1^2}
-\frac{k}{2k_1}\ln\frac{k+2k_1}{k-2k_1}\right)
+\cdots\right]
\nonumber\\
&&+\frac{g^2N_c}{8\pi^2}\int_0^\infty dk_1 \, k_1 n_B(k_1)\left[8
-\frac{4k^2}{k_1^2}
+\left(\frac{8k_1}{k}-\frac{4k}{k_1}+\frac{k^3}{k_1^3}\right)
\ln\frac{k+2k_1}{k-2k_1}
\right.\nonumber\\
&&\qquad\left.
-\frac{\omega^2}{k^2}\left(
\frac{3k^6-8k^4k_1^2-112k^2k_1^4+256k_1^6}{k_1^2(k^2-4k_1^2)^2}
+\frac{5k(k^2-k_1^2)^2}{2k_1^5}
\ln\frac{k+k_1}{|k-k_1|}
\right.\right.\nonumber\\
&&\qquad\left.\left.
-\frac{k(5k^4-12k^2k_1^2+24k_1^4)}{4k_1^5}
\ln\frac{k+2k_1}{k-2k_1}\right)
+\cdots
\right] \, ,
\end{eqnarray}
where the first and second integrals correspond to quark and gluon 
contributions, respectively. 
The $\mathcal{O}(\omega^0)$ and $\mathcal{O}(\omega^2)$ terms yield the rest 
mass shift and the relative correction, respectively, 
while the transverse part is subleading.  
If the momentum transfer is smaller than the incoming parton momentum, we have  
\begin{eqnarray}
{\rm Re}\, \Pi_L(K)&\sim&
\frac{2g^2N_f}{\pi^2}\int_0^\infty dk_1 \, k_1n_F(k_1)
\left(1-\frac{\omega^2}{k^2}\right)
\nonumber\\
&&\qquad+\frac{2g^2N_c}{\pi^2}\int_0^\infty dk_1 \, k_1 n_B(k_1)
\left(1-\frac{\omega^2}{k^2}\right)
+\mathcal{O}\left(\frac{k^2}{k_1^2},\omega^4\right) \, ,
\end{eqnarray}
and the thermal mass shift is estimated to be 
$\delta m_{\rm rest}(1-\frac{v^2}{3})$. 
However, we need to use the resummed propagators when $k\sim gT$.

\section{Damping rate}
\label{damp}

The static potential for a heavy quark-antiquark pair has been computed at 
high temperature, $T\gtrsim \frac{1}{r}\sim mg^2$ 
\cite{Laine:2006ns,Brambilla:2008cx}. 
Employing the potential, the quarkonium contribution 
to the spectral function of the electromagnetic current has been 
numerically computed \cite{Laine:2007gj,Burnier:2007qm}.  
Especially for $T\gg \frac{1}{r}\sim m_D$, the real part $\mathcal{O}(g^2m_D)$ 
of the potential is smaller than the imaginary part $\mathcal{O}(g^2T)$ 
(whose $m_Dr\ll 1$ expansion yields $\Gamma\sim g^2T(m_Dr)^2$ discussed 
in Section \ref{inelscatt}), so the heavy quark-antiquark pair is expected to 
decay before forming a bound state \cite{Brambilla:2008cx}. 
It has been noticed that the real part correction coincides with the free 
energy of static heavy quark-antiquark \cite{Gava:1981qd,Petreczky:2005bd} 
(or twice the rest mass shift of heavy quark), 
while the imaginary part approaches twice the heavy quark damping rate 
\cite{Pisarski:1988vd,Pisarski:1993rf} (with minus sign) 
in the large $r$ limit \cite{Beraudo:2007ky,Brambilla:2008cx}. 
The heavy quark propagator can be approximated by its nonrelativistic particle 
pole:
\begin{equation}
\label{covver}
D_b\left(P\pm \frac{Q+K}{2}\right)\rightarrow
\frac{\pm i(1\pm\slashed{V})}{2
(\pm p_0+ \frac{\omega- \k\cdot\v}{2}-\frac{E}{2}-\frac{\p^2}{2m}+
i\frac{\Gamma_b}{2})} \, ,
\end{equation}  
where $V=\frac{1}{\sqrt{1-v^2}}(1,\v)$ and $\Gamma_b$ is the heavy quark width. 
For a heavy quark-antiquark pair with the effective 
vertex $V_{\rm eff}^{\mu\nu}(K)$ in Fig. \ref{YselfE}, 
\begin{multline}
\int \frac{dp^0}{2\pi}
D_b\left(P+\frac{Q+K}{2}\right)
V_{\rm eff}^{\mu\nu}(K)
D_b\left(P-\frac{Q+K}{2}\right)
V_{\rm eff}^{\rho\sigma \dagger}(K)
\\ \simeq
\frac{1+\slashed{V}}{2}
V_{\rm eff}^{\mu\nu}(K)
\frac{1-\slashed{V}}{2}
V_{\rm eff}^{\rho\sigma \dagger}(K)
\frac{i}{\omega-\k\cdot\v-E-\frac{\p^2}{m}+i\Gamma_b} \, ,
\label{loop}
\end{multline}
where the width of the quarkonium state is twice that of heavy quark 
\cite{Strassler:1990nw}.

In this work, we are interested in the kinematical regime where 
$\frac{1}{r}\gg T$ and $m_D \gg E$ are satisfied so that 
a dipole approximation is appropriate and quarkonium dissociation can be 
parallel to heavy quark diffusion.  
In Eq. (\ref{loop}), without $\Gamma_b$ we get $\delta(\omega-\k\cdot\v)$ 
which appears throughout the paper.  
We have calculated the thermal width and mass shift of quarkonium moving in 
quark-gluon plasmas in the previous two sections. 
They provide the thermal corrections, $\delta m-i\frac{\Gamma}{2}$, to the 
singlet static potential \cite{Brambilla:2008cx,Brambilla:2016wgg}. 
In the following, we employ the complex potential to estimate the spectral 
density of nonrelativistic quarkonium near the threshold 
$\omega\sim 2m+\frac{\q^2}{4m}$. 
The result is then used to compute the quarkonium damping rate which is 
comparable with the thermal width.

As discussed in Refs. \cite{Laine:2006ns,Laine:2007gj,Burnier:2007qm}, 
a heavy quark current-current correlator 
$C^>(X)=\langle \hat{J}^\mu(X) \hat{J}_\mu(0)\rangle$, where 
$\hat{J}^\mu(X)=\hat{\bar{\psi}}(X)\gamma^\mu\hat{\psi}(X)$, 
satisfies a Schr\"{o}dinger-type equation with a complex potential:  
\begin{equation}
\left[i\frac{\partial}{\partial t}
-\left(2m+\frac{q^2}{4m}-\frac{\nabla_{\r}^2}{m}+V(r)\right)\right] 
C^>(t,r)=0 \, .
\end{equation}
Although one can solve the equation numerically, we consider a simple 
case in the static limit: 
${\rm Re}\,V(r)\approx-\frac{g^2C_F}{4\pi r}e^{-m_Dr}$ 
(plus the thermal mass shift) and 
$\int d^3\r\, {\rm Im}\,V(r)=-\frac{\Gamma}{2}$ at short distance $r$ 
(we assume that only the real part depends on $r$). 
Since the potential is independent of time, 
$C^>(t,r)\propto e^{-i(E_{\q}-i\frac{\Gamma}{2}) t}$, where 
$E_{\q}=2m+\frac{\q^2}{4m}-E$. 
The radial part is given by a bound state for a Debye-screened Coulombic 
potential, 
\begin{equation}
\left[\frac{\nabla_{\r}^2}{m}-{\rm Re}\, V(r)\right] 
\psi(r)=E\psi(r) \, .
\end{equation}
In the limit $r\rightarrow0$, we take a Fourier transform with respect to 
time:   
\begin{eqnarray}
\rho(\omega)
\propto \int_0^{\infty} dt \, 
e^{-\frac{\Gamma}{2}t}[\cos(E_{\q} t)\cos(\omega t)+
\sin(E_{\q}t)\sin(\omega t)]
=\frac{\frac{\Gamma}{2}}{(\omega-E_{\q})^2+\frac{\Gamma^2}{4}}
\, .
\label{Yspec}
\end{eqnarray}
Retaining the leading terms in the real and ($\r$-integrated) imaginary parts 
of the potential, 
we obtain a Breit-Wigner form of the spectral density for a 
spherically symmetric wave function: the imaginary part of the potential 
provides a finite width of a bound state.

The damping rate is given by the imaginary part of the self-energy at 
the pole (which has an imaginary part) in the propagator, while on the 
mass-shell half the interaction rate is expected \cite{Bellac:2011kqa}. 
Since the spectral density has been estimated in Eq. (\ref{Yspec}), 
we follow Ref. \cite{Pisarski:1993rf} for heavy quark to compute 
the damping rate of nonrelativistic quarkonium. 
After integrating a wave function and averaging over polarization, 
\begin{multline}
\frac{\Gamma}{2}=\frac{4g^2a_0^2T}{3}
\int\frac{d^3\k}{(2\pi)^3}\int \frac{d\omega}{\omega} \int d\omega_\Upsilon
\frac{\frac{\Gamma}{2}}{[(\omega_\Upsilon-E_{\q-\k})^2+\frac{\Gamma^2}{4}]}
\\
\times
\left[k^2\rho_L(K)+2\omega^2\rho_T(K)\right]
\delta(\omega+\omega_\Upsilon-E_{\q}) \, .
\end{multline}
We redefine $\omega_\Upsilon\rightarrow\omega_\Upsilon+E_{\q-\k}$ and 
$E_{\q-\k}\approx m_\Upsilon+\frac{\q^2}{2m_\Upsilon}-\k\cdot\v$. 
Then the delta function becomes $\delta(\omega+\omega_\Upsilon-\k\cdot\v)$. 
Since $\omega_\Upsilon\sim g^4T^3a_0^2$ is much smaller than the scale of 
$\omega$ or $vk$, we can perform the integrations over $\omega_\Upsilon$ and 
$\omega$ separately: 
$\int_{-vk}^{vk}d\omega_\Upsilon
\frac{\frac{\Gamma}{2}}{(\omega_\Upsilon^2+\frac{\Gamma^2}{4})}
=2\,\arctan\left(\frac{2vk}{\Gamma}\right)$. 
Finally, approximating the spectral densities in the limit 
$\omega\rightarrow 0$, 
\begin{eqnarray}
\rho_L(K)&\simeq&-{\rm Im}\, \frac{1}{\pi\big[k^2+
m_D^2\big(1+i\frac{\pi\omega}{2k}\big)\big]} \, ,
\nonumber\\
\rho_T(K)&\simeq&{\rm Im} \,\frac{1}{\pi\big(k^2-i\frac{m_D^2\pi\omega}{4k}\big)} \, ,
\end{eqnarray}
we have 
\begin{equation}
\frac{\Gamma}{2}\simeq\frac{g^2a_0^2}{3\pi}\left(1+\frac{v^2}{3}\right)
m_D^2T\ln\frac{T}{m_D} \, .
\end{equation}
This result agrees with the thermal width, Eq. (\ref{G1S}), 
for small $v$.

Due to the quarkonium-gluon effective vertex involving momentum and energy 
transfer, the long-range interactions in both longitudinal and transverse 
gluon parts are screened by the Debye mass in the leading-log damping rate. 
This is in contrast to the heavy quark damping rate where the 
transverse part is in need of an infrared regulator such as the magnetic mass 
($\sim g^2T$) or the imaginary part of the pole (that is, the damping rate 
itself) \cite{Thoma:1990fm,Pisarski:1993rf}. 
The damping rate of slow heavy quark has been 
computed for $g^{\frac{2}{3}}\ll v\ll 1$: 
$\gamma_{HQ}
=\frac{g^2C_FT}{8\pi}\Big(1+\frac{v}{2}
\ln\frac{\pi m_D^2 v^3}{4\gamma_{HQ}^2}\Big)$, 
where $\gamma_{HQ}$ inside the logarithm is replaced by its value at $v=0$ 
\cite{Pisarski:1993rf}.

\section{Summary}
\label{summary}

We have presented the velocity dependence of the thermal width and mass shift 
for heavy quarkonia slowly moving in hot QCD plasmas. 
In the kinematical regime where $\frac{1}{r}\gg T$ and $m_D \gg E$, 
the longitudinal gluon contribution is same for both the thermal width and 
the momentum diffusion coefficient of heavy quarks, but the transverse gluon 
part is different. 
For nonrelativistic quarkonia, we have discussed the thermal width in 
relation to the damping rate which is proportional to the imaginary part of 
the color-singlet self-energy. 
We have also discussed the thermal corrections to the dispersion relation, 
the effective rest mass shift and its correction (kinetic mass shift) for 
small momentum transfer. 
On the other hand, at a higher temperature regime $T\gg \frac{1}{r}$ 
a heavy quark-antiquark pair is likely to decay before forming a bound state. 
The real and imaginary part of the static potential approach 
the mass shift and the damping rate, respectively, of heavy quark and 
antiquark in the large $r$ limit.

For a slowly moving bound state, the thermal width agrees with the momentum 
diffusion coefficient of heavy quark up to $\mathcal{O}(v^2)$. 
However, the thermal width of nonrelativistic quarkonium is same as neither 
the heavy quark diffusion coefficient nor the damping rate of heavy 
quark-antiquark pair in the short $r$ limit. 
The relative correction to the rest mass shift for quarkonium is also 
different from that for heavy quark. 
In studying quarkonium dynamics, the transport coefficients of heavy quarks 
and quarkonia might be relevant in different kinematical regimes. 
Eventually the nonperturbative determination of the transport 
coefficients is desirable, while our perturbative analysis of the velocity 
dependence can supplement current lattice QCD computations.

\appendix

\section{Quarkonium-gluon interaction}
\label{gluodiss}

In this appendix, we provide the details on the leading-order 
quarkonium-gluon interaction which has been derived with the Bethe-Salpeter 
amplitude \cite{Oh:2001rm} through dipole interaction \cite{Bhanot:1979vb}. 
The Bethe-Salpeter equation for a bound state is \cite{Salpeter:1951sz}
\begin{multline}
\label{BSeq}
\Gamma_{\rm BS}^\mu(P_1,-P_2)
\\
=ig^2C_F\int\frac{d^4K}{(2\pi)^4}V(K)\gamma^\nu 
D_b(P_1+K)\Gamma_{\rm BS}^\mu(P_1+K,-P_2+K)D_b(-P_2+K)\gamma_\nu \, ,
\end{multline}
where $V(K)\simeq-\frac{1}{\k^2}$ after computing a dominant residue. 
For $\q,\p_{1,2}\gg \k$ and in the rest frame of quarkonium, the 
Bethe-Salpeter amplitude reduces to 
\begin{equation}
\label{BSamp}
\Gamma_{\rm BS}^\mu\left(P+\frac{Q}{2},P-\frac{Q}{2}\right)
=\sqrt{\frac{m_\Upsilon}{N_c}}\left(E+\frac{\p^2}{m}\right)\psi(\p)
\frac{1+\gamma^0}{2}\gamma^i\delta^{\mu i}\frac{1-\gamma^0}{2} \, ,
\end{equation}
and Eq. (\ref{BSeq}) becomes the nonrelativistic Schr\"{o}dinger equation for 
a Coulombic bound state, 
\begin{equation}
\left(E+\frac{\p^2}{m}\right)\psi(\p)=g^2C_F\int \frac{d^3\k}{(2\pi)^3}V(\k)
\psi(\p+\k) \, .
\end{equation}
As in Eq. (\ref{covver}), the leading expression in the plasma rest frame is 
expected with the replacement, $1\pm\gamma^0\rightarrow 1\pm\slashed{V}$.

Inelastic parton scattering is a 
dominant dissociation process of a bound state like $\Upsilon(1S)$ which 
survives in high-temperature plasmas, whereas  
at low temperature near the phase transition gluo-dissociation 
($\Upsilon+g\rightarrow b+\bar{b}$) becomes effective 
\cite{Hong:2018vgp}. 
Using Eq. (\ref{BSamp}), the scattering amplitude of gluon absorption 
has been obtained as follows \cite{Oh:2001rm}: 
\begin{eqnarray}
\mathcal{M}_{\rm gluo}^{\mu\nu}
&=&-g\sqrt{\frac{m_\Upsilon}{N_c}}\left[\k\cdot\frac{\partial
\psi(\p)}{\partial\p}\delta^{\mu 0}+k_0\frac{\partial\psi(\p)}{\partial p^i}
\delta^{\mu i}\right]\delta^{\nu j}\bar{u}(P_1)\frac{1+\gamma^0}{2}\gamma^j
\frac{1-\gamma^0}{2}T^av(P_2) \, ,
\nonumber\\
&\equiv&\bar{u}(P_1)V_{\rm eff}^{\mu\nu}(K)v(P_2) \, ,
\label{effver}
\end{eqnarray}
where a quarkonium-gluon effective vertex $V_{\rm eff}^{\mu\nu}(K)$ 
can be defined. 
The effective vertex is similar to the pNRQCD vertex \cite{Brambilla:2004jw}, 
with $\frac{\partial\psi(\p)}{\partial\p}$ corresponding to $\r$ multiplied by 
a bound state.

The thermal width by gluo-dissociation is  
\begin{multline}
\Gamma_{\rm gluo}
=\frac{1}{2d_\Upsilon q^0}
\int\frac{d^3\k}{(2\pi)^32k^0}
\int\frac{d^3\p_1}{(2\pi)^32p_{10}}
\int\frac{d^3\p_2}{(2\pi)^32p_{20}} \\
\times (2\pi)^4\delta^4(Q+K-P_1-P_2)
|\mathcal{M}|^2_{\rm gluo}n(k) \, ,
\end{multline}
which corresponds to a different imaginary part of the heavy quark-antiquark 
potential from inelastic parton scattering \cite{Brambilla:2008cx}. 
Gluo-dissociation is dominant in the kinematical regime 
$\frac{1}{r}\gg T\gg E\gg m_D$ \cite{Brambilla:2013dpa} and 
is not directly related to the momentum diffusion coefficient of heavy quark. 
In contrast to inelastic parton scattering, the thermal width by 
gluo-dissociation decreases with $v$ \cite{Hong:2019ade}.

\section*{Acknowledgments}

This research was supported by Basic Science Research Program through the
National Research Foundation of Korea (NRF) funded by the Ministry of Education 
(No. 2021R1I1A1A01054927).


\begin{thebibliography}{99}

\bibitem{Matsui:1986dk}
T.~Matsui and H.~Satz,
Phys. Lett. B \textbf{178}, 416-422 (1986).

\bibitem{Laine:2006ns}
M.~Laine, O.~Philipsen, P.~Romatschke and M.~Tassler,
JHEP \textbf{03}, 054 (2007)
[arXiv:hep-ph/0611300 [hep-ph]].

\bibitem{Brambilla:2008cx}
N.~Brambilla, J.~Ghiglieri, A.~Vairo and P.~Petreczky,
Phys. Rev. D \textbf{78}, 014017 (2008)
[arXiv:0804.0993 [hep-ph]].

\bibitem{Hong:2018vgp}
J.~Hong and S.~H.~Lee,
Phys. Rev. C \textbf{99}, 034905 (2019)
[arXiv:1811.07607 [nucl-th]].

\bibitem{Brambilla:2013dpa}
N.~Brambilla, M.~A.~Escobedo, J.~Ghiglieri and A.~Vairo,
JHEP \textbf{05}, 130 (2013)
[arXiv:1303.6097 [hep-ph]].

\bibitem{Brambilla:2016wgg}
N.~Brambilla, M.~A.~Escobedo, J.~Soto and A.~Vairo,
Phys. Rev. D \textbf{96}, 034021 (2017)
[arXiv:1612.07248 [hep-ph]].

\bibitem{Brambilla:2019tpt}
N.~Brambilla, M.~A.~Escobedo, A.~Vairo and P.~Vander Griend,
Phys. Rev. D \textbf{100}, 054025 (2019)
[arXiv:1903.08063 [hep-ph]].

\bibitem{Hong:2019ade}
J.~Hong and S.~H.~Lee,
Phys. Lett. B \textbf{801}, 135147 (2020)
[arXiv:1909.07696 [nucl-th]].

\bibitem{Escobedo:2013tca}
M.~A.~Escobedo, F.~Giannuzzi, M.~Mannarelli and J.~Soto,
Phys. Rev. D \textbf{87}, 114005 (2013)
[arXiv:1304.4087 [hep-ph]].

\bibitem{Weldon:1982aq}
H.~A.~Weldon,
Phys. Rev. D \textbf{26}, 1394 (1982).

\bibitem{Chu:1988wh}
M.~C.~Chu and T.~Matsui,
Phys. Rev. D \textbf{39}, 1892 (1989).

\bibitem{Oh:2001rm}
Y.~s.~Oh, S.~Kim and S.~H.~Lee,
Phys. Rev. C \textbf{65}, 067901 (2002)
[arXiv:hep-ph/0111132 [hep-ph]].

\bibitem{Bhanot:1979vb}
G.~Bhanot and M.~E.~Peskin,
Nucl. Phys. B \textbf{156}, 391 (1979).

\bibitem{Moore:2004tg}
G.~D.~Moore and D.~Teaney,
Phys. Rev. C \textbf{71}, 064904 (2005)
[arXiv:hep-ph/0412346 [hep-ph]].

\bibitem{Braaten:1989mz}
E.~Braaten and R.~D.~Pisarski,
Nucl. Phys. B \textbf{337}, 569 (1990).

\bibitem{Weldon:1982bn}
H.~A.~Weldon,
Phys. Rev. D \textbf{26}, 2789 (1982).

\bibitem{Braaten:1991dd}
E.~Braaten and T.~C.~Yuan,
Phys. Rev. Lett. \textbf{66}, 2183 (1991).

\bibitem{Braaten:1991jj}
E.~Braaten and M.~H.~Thoma,
Phys. Rev. D \textbf{44}, 1298 (1991).

\bibitem{Braaten:1991we}
E.~Braaten and M.~H.~Thoma,
Phys. Rev. D \textbf{44}, R2625 (1991).

\bibitem{Weldon:1983jn}
H.~A.~Weldon,
Phys. Rev. D \textbf{28}, 2007 (1983).

\bibitem{Caron-Huot:2007rwy}
S.~Caron-Huot and G.~D.~Moore,
Phys. Rev. Lett. \textbf{100}, 052301 (2008)
[arXiv:0708.4232 [hep-ph]].

\bibitem{Chesler:2009yg}
P.~M.~Chesler, A.~Gynther and A.~Vuorinen,
JHEP \textbf{09}, 003 (2009)
[arXiv:0906.3052 [hep-ph]].

\bibitem{Kajantie:1982xx}
K.~Kajantie and J.~I.~Kapusta,
Annals Phys. \textbf{160}, 477 (1985).

\bibitem{Heinz:1986kz}
U.~W.~Heinz, K.~Kajantie and T.~Toimela,
Annals Phys. \textbf{176}, 218 (1987).

\bibitem{Kapusta:2006pm}
J.~I.~Kapusta and C.~Gale,
``Finite-temperature field theory: Principles and applications''
(Cambridge University Press, 2006).

\bibitem{Laine:2007gj}
M.~Laine,
JHEP \textbf{05}, 028 (2007)
[arXiv:0704.1720 [hep-ph]].

\bibitem{Burnier:2007qm}
Y.~Burnier, M.~Laine and M.~Vepsalainen,
JHEP \textbf{01}, 043 (2008)
[arXiv:0711.1743 [hep-ph]].

\bibitem{Gava:1981qd}
E.~Gava and R.~Jengo,
Phys. Lett. B \textbf{105}, 285 (1981).

\bibitem{Petreczky:2005bd}
P.~Petreczky,
Eur. Phys. J. C \textbf{43}, 51 (2005)
[arXiv:hep-lat/0502008 [hep-lat]].

\bibitem{Pisarski:1988vd}
R.~D.~Pisarski,
Phys. Rev. Lett. \textbf{63}, 1129 (1989).

\bibitem{Pisarski:1993rf}
R.~D.~Pisarski,
Phys. Rev. D \textbf{47}, 5589 (1993).

\bibitem{Beraudo:2007ky}
A.~Beraudo, J.~P.~Blaizot and C.~Ratti,
Nucl. Phys. A \textbf{806}, 312 (2008)
[arXiv:0712.4394 [nucl-th]].

\bibitem{Strassler:1990nw}
M.~J.~Strassler and M.~E.~Peskin,
Phys. Rev. D \textbf{43}, 1500 (1991).

\bibitem{Bellac:2011kqa}
M.~L.~Bellac,
``Thermal Field Theory''
(Cambridge University Press, 1996).

\bibitem{Thoma:1990fm}
M.~H.~Thoma and M.~Gyulassy,
Nucl. Phys. B \textbf{351}, 491 (1991).

\bibitem{Salpeter:1951sz}
E.~E.~Salpeter and H.~A.~Bethe,
Phys. Rev. \textbf{84}, 1232 (1951).

\bibitem{Brambilla:2004jw}
N.~Brambilla, A.~Pineda, J.~Soto and A.~Vairo,
Rev. Mod. Phys. \textbf{77}, 1423 (2005)
[arXiv:hep-ph/0410047 [hep-ph]].







\end{thebibliography}
\end{document}